\documentclass[10pt,twocolumn]{article}
\usepackage[margin=1in]{geometry}
\usepackage{amsmath}
\usepackage{abstract}
\usepackage{amsfonts,color}
\usepackage{amssymb}
\usepackage{graphicx}
\usepackage{hyperref}
\begin{document}
\title{\bf{Bondi flow revisited}}
\author
{Satadal Datta\\
Harish-Chandra research Institute, Chhatnag Road, Jhunsi, Allahabad-211019, INDIA\\
Email: satadaldatta@hri.res.in}
\twocolumn[
\maketitle
\begin{onecolabstract}
Newtonian spherically symmetric transonic 
accretion has been studied by including 
the mass of the accreting matter, while considering the growth
of the accretor itself to be negligibly small. A novel iterative method 
has been introduced to accomplish that task. It has been 
demonstrated that the inclusion of the mass of the fluid changes the 
critical properties of the flow as well as the topological phase 
portraits of the stationary integral solution. 
\end{onecolabstract}
]
\vspace*{0.33cm}
\section{Introduction}
Spherically symmetric Bondi \cite{a} flow studies the dynamics 
of the infalling test fluids. In the present work, we introduce a novel 
iterative method to study the effect of the inclusion of the mass
of the accreting material. We did not consider the growth of the accretor
as a consequence of the accretion and hence the direct effect of the 
inclusion of the self gravity has not been studied. For usual astrophysical 
accretion, the aforementioned approximation, that the accretion rate 
as well as the time scale to study the problem is not so large that the 
mass of the accretor will change -- is a valid approximation. 

We consider steady state accretion of one temperature fluid onto 
a non spinning stationary (the observer is in the co-moving frame)
accretor. We consider a more appropriate effective gravitational 
potential in comparison to what had been assumed by Chia \cite{b}. 

We found that the inclusion of fluid mass alters the location of the 
critical point of the flow as well as the values of the values of 
the accretion variables measured at the critical point. We also find that 
the Mach number vs radial distance profile, the usual topology of the 
phase portrait of the Bondi flow, takes a different form for such massive 
accretion. We characterize our flow profile using a realistic set of 
astrophysically relevant parameters. \\
\\
\section{Governing Equations}
As we are considering steady flow, the Bernoulli's equation reads as \cite{d}\footnote{In this paper, the basic fluid equations are taken from this cited book.}

\begin{equation}
\frac{u^{2}}{2}+\int \frac{dp}{\rho}+V(r)=constant
\end{equation}
where fluid velocity is u, pressure is p and V(r) is the gravitational potential due to the accretor and the infalling fluid itself. V(r) satisfying Poisson's equation is as below,
\begin{equation}
V(r)=-\frac{GM}{r}-\frac{4\pi G}{r}\int_{R_\star}^{r}\rho r^{2}dr-4\pi G\int_{r}^{\infty}\rho(r)rdr
\end{equation}
where R$_\star$ is the radius of the accretor.
For brevity we are considering adiabatic flow.
So our baratropic equation is
\begin{equation}
p=K\rho^{\gamma}
\end{equation}
where K is a constant. And sound speed follows the relation,
\begin{equation}
c_{s}^{2}=\left(\frac{\partial p}{\partial\rho}\right)_{s}=\frac{\gamma p}{\rho}
\end{equation}
 where s = entropy. Now, 
 we have the energy of fluid element per unit mass is E. Now, at infinity fluid velocity is zero so according to Bernoulli's equation,
\begin{equation}
\frac{u^{2}}{2}+\frac{c_{s}^{2}}{(\gamma-1)}+V(r)=E=\frac{c_{s\infty}^{2}}{(\gamma-1)}
\end{equation}
Physically, we are taking our infinity to be such a large distance such that the potential terms are vanishing there. We have discussed this issue in details later in section 6.
Now we define certain dimensionless variables as 
\begin{align*}
x=\frac{r}{\frac{GM}{c_{s\infty}^{2}}},
y=\frac{u}{c_{s\infty}}
,z=\frac{\rho}{\rho _{\infty}}
\end{align*}
where $c_{s\infty}$ is the sound speed at infinity and $\rho _{\infty}$ is the density of fluid at infinity.
 From equation (5), introducing the dimension less variables, we get
\begin{equation}
\frac{y^2}{2}+\frac{z^{(\gamma-1)}}{(\gamma-1)}-\frac{1}{x}-\alpha\left(\frac{I_{1}}{x}+I_{2}\right)=0
\end{equation}
where $\alpha=4\pi\rho_{\infty}G^3M^2c_{s\infty}^{-6}$,
$I_{1}=\int\limits_{x*}^{x}zx^2dx$ and
$I_{2}=\int\limits_{x}^{\infty}zxdx$ with $x*=\frac{R_\star}{(\frac{GM}{c_{s\infty}^{2}})}$\\

Thus the mass accretion rate which is obtained by integrating the continuity equation,
\begin{align*}
\dot{M}=4\pi r^2 \rho u=4\pi(GM)^2 c_{s\infty}^{-3}\rho_{\infty}x^2yz
\end{align*}
We define another dimensionless variable $\lambda=x^2yz$ and from above equation mass accretion rate is proportional to $\lambda$. Now, we can rewrite equation (6) as folows,
\begin{equation}
\frac{\lambda^2}{2z^2}+\frac{(z^{\gamma-1}-1)}{(\gamma-1)}x^4-(1+\alpha I_{1})x^3-\alpha x^4I_{2}=0
\end{equation}
\section{Calculations of Critical Point of the flow}
Now, our aim is to maximize accretion rate. We will find x, y, z for which $\lambda$ is maximized. So, from equation (7) setting $\frac{\partial\lambda}{\partial x}\mid_{(x=xc,y=yc,z=zc)}=0$ and $\frac{\partial\lambda}{\partial z}\mid_{(x=xc,y=yc,z=zc)}=0$, we get respectively
\begin{equation}
\frac{4(z_{c}^{(\gamma -1)}-1)x_{c}}{(\gamma -1)}-3(1+\alpha I_{1c})-4\alpha x_c I_{2c}=0
\end{equation}
and
\begin{equation}
\lambda_{c}^{2}=z_{c}^{\gamma +1}x_{c}^4
\end{equation}
Using equation (9), from equation (7), we get
\begin{equation}
\frac{z_{c}^{(\gamma -1)}x_{c}}{2}+\frac{(z_{c}^{(\gamma -1)}-1)x_{c}}{(\gamma -1)}-1-\alpha (I_{1c}+\alpha x_c I_{2c})=0
\end{equation}
As a consequence we have the following:\\
The critical values x$_{c}$, y$_{c}$ and z$_{c}$ correspond to the Mach number to be unity, i.e; the transonic curve corresponds to maximum accretion rate.
\\
We have from definition
\begin{align*}
\lambda_{c}=x_{c}^2y_cz_c
\end{align*}
Using equation (9), we get,\\
\\
Mach number = $\textit{m}=y_cz_c^{-\frac{(\gamma -1)}{2}}=1$\\
-------------------------------------------------
\\
Now, from equation (8) and (10), we get,
\\
\begin{equation}
\frac{x_c}{x_b}=\frac{1+\alpha I_{1c}}{1+\alpha (\gamma-1)I_{2c}}
\end{equation}
\begin{equation}
(\frac{z_c}{z_b})^{(\gamma-1)}=(1+\alpha (\gamma-1)I_{2c})
\end{equation}
Using equation (9), we get
\begin{equation}
\frac{\lambda_c}{\lambda_b}=(1+\alpha I_{1c})^2(1+\alpha (\gamma-1)I_{2c})^{\frac{(5-3\gamma)}{2(\gamma-1)}}
\end{equation}
where $x_b,~z_b,~\lambda_b$ are the critical values in case of spherical usual Bondi accretion without considering fluid mass. These values are respectively as follows:
\begin{align*}
x_b=\frac{(5-3\gamma)}{4}\\
z_b^{(\gamma-1)}=\frac{1}{2x_b}\\
\lambda_b^2=z_b^{(\gamma+1)}x_b^4
\end{align*}
\section{Equation of motion of the infalling fluid}
To find the behaviour of the Mach number vs radial distance profile for self gravitating steady state Bondi accretion, we have to know the equation of motion of the infalling fluid. According to Euler's momentum conservation equation:
\begin{align*}
\rho\frac{D\vec{u}}{Dt}=-\vec{\triangledown}p-\rho\vec{\triangledown}V
\end{align*}
V is the external potential, in our case V is the gravitational potential. So considering steady flow, we have
\begin{equation}
u\frac{du}{dr}=-\frac{1}{\rho}\frac{dp}{dr}-\frac{GM}{r^2}-\frac{4\pi G}{r^2}\int_{R_\star}^{r}\rho r^2dr
\end{equation} Again considering spherically symmetric steady flow, we have from continuity equation,
accretion rate, \begin{equation}
\dot{M}=4\pi r^2\rho u=constant
\end{equation}
Considering steady flow, using equation (3), (4), (14), we have
\begin{equation}
\frac{du}{dr}=\frac{(\frac{2c_s^2}{r}-\frac{GM}{r^2}-\frac{4\pi G}{r^2}\int_{R_\star}^{r}\rho r^2dr)}{(u-\frac{c_s^2}{u})}
\end{equation}
Now, using equation (15), (3), (4), we have 
\begin{equation}
\frac{dc_s}{dr}=\frac{c_s(1-\gamma)}{2}\left(\frac{2}{r}+\frac{1}{u}\frac{du}{dr}\right)
\end{equation}
If we can solve these equations we can have the Mach number vs radial distance profile except at the transonic point because at transonic point from equation (16) takes 0/0 form comes. So, applying L'Hospital's rule there we have a quadratic equation as
\begin{align*}
Aq^2+Bq+C=0
\end{align*}
where,
\begin{align*}
& A=(1+\gamma)\\
& B=\frac{4c_{sc}(\gamma-1)}{r_c}\\
& C=\frac{2c_{sc}^2(2\gamma-1)}{r_c^2}-\frac{2GM}{r_c^3}+4\pi G\rho_c-\frac{8\pi G}{r_c^3}I_{1c}\\
& q=\frac{du}{dr}|_{critical~point}\\
\end{align*} 
$c_{sc}$ is the sound speed at critical point. $r_c$ is the radial distance of critical point from the centre of the accretor. So,
\begin{equation}
q=\frac{-B\pm\sqrt{B^2-4AC}}{2A}
\end{equation}
Still, the problem is unsolvable because to find the values of the dimensionless variables at critical point (which will serve as the initial conditions to solve differential equations (16) and (17)), we need density profile of the infalling fluid in the first place and we have no idea about the density profile. So in the next section, we develop a methodology to solve the differential equations (16) and (17).
\section{Method of iteration}
In equation (11), (12), (13) and (16) $\rho$ appears to be a variable about which we have no information. So, we will consider that even if we include mass of the medium it does not change the critical values $x_c, ~y_c,~ z_c~$and$~\lambda_c$ abruptly. According to equations (11), (12) and (13), in this case, the value of $\alpha$ is small. Now we will find the values of the dimensionless quantities from equations (11), (12) and (13) by method of iteration. To find the critical parameters from the above equations we have to find the integrals $I_{1c}$ and $I_{2c}$. So, as our first iteration we will put the value of z as the usual z of spherically symmetric Bondi flow. x$_c$ appears in the definite integrals $I_{1c}$ and $I_{2c}$ as upper limit and lower limit respectively and in this case as our 1st iteration, we will replace x$_c$ by x$_b(=(5-3\gamma)/4)$. Thus as our 1st iteration, we can find the values of the integrals. And then we can compute the critical parameters from equation (11), (12) and (13). Again, as our 2nd iteration, we can put these new values as the limits of the integrals described before and so on and thus we can improve our precision in each iteration. As we are considering weak gravitational field due to fluid mass, we have described only the 1st iteration here. In the 2nd iteration the terms of order $\alpha^2$ will appear (obvious from equation (11), (12) and (13)). We are neglecting the order $\alpha ^2$ and the terms of higher order in $\alpha$.  So, in the 1st iteration, the values of $I_{1c}$ and $I_{2c}$ are as below.
\begin{equation}
I_{1c}=\int_{x*}^{x_b}\rm z_{Bondi}(x)x^2dx
\end{equation}
\begin{equation}
I_{2c}=\int_{x_b}^{\infty}\rm z_{Bondi}(x)xdx
\end{equation}
where $\rm z_{Bondi}(x)$ is z for the usual Bondi flow without considering fluid mass. Now, again we have a problem that there is no proper analytic form for z$\rm_{Bondi}(x)$, so we will find the integrals using the numerical values of z$\rm_{Bondi}(x)$ at different x . 
\section{The limits of the integrals}
x$_b$ is the upper limit and lower limit of the integrals $I_{1c}$ and $I_{2c}$ respectively. The lower limit appearing in the integral $I_{1c}$ is x*. Now, in the integral, we can put x* using values of radius of some known stars and we can put the mass of the stars in equations. Alternatively, we can find the radius of a star when the mass of a star is given from any empirical relation. We have used the following emperical relations\cite{c} between mass and radius of a Main Sequence star.
\begin{equation}
\frac{R_\star}{R_\odot}=1.06\left(\frac{M}{M_\odot}\right)^{0.945}~~~~~~~M<1.66M_\odot
\end{equation}
and
\begin{equation}
\frac{R_\star}{R_\odot}=1.33\left(\frac{M}{M_\odot}\right)^{0.555}~~~~~~~M>1.66M_\odot
\end{equation}
where R$_\star$, R$_\odot$ are the radius of the star and of the Sun and M, M$_\odot$ are the mass of the star and of the Sun respectively. Thus the lower limit of the integral $I_{1c}$ is fixed.\\
 Now, $\infty$ appears in the upper limit of the integral $I_{2c}$. By $\infty$ we mean some large distance from the accretor where gravity is weak. Let's call this large distance to be r$_\infty$ and corresponding value of the dimensionlesss parameter to be x$_\infty$. The mass, the energy and $\gamma$ of the fluid are given. At infinity as the velocity of the fluid is zero, so we have
 \begin{equation}
\frac{c_{s_\infty}^{2}}{(\gamma-1)}-\frac{GM}{r_\infty}-\frac{4\pi G}{r_\infty}\int_{R_\star}^{r_\infty}\rho r^{2}dr=E
 \end{equation}
 Now, we will compare the magnitude of the 1st term in potential with energy and we will set r$_\infty$ to some value such that the term $\frac{GM}{r_\infty}$ is negligibly small compared to energy of the infalling fluid.\\
 Quantitatively, we set a very small quantity Q1 according to our desired precision as below.
 \begin{equation}
 Q1=\frac{(\frac{GM}{r_\infty})}{E}
 \end{equation}
 Now, we can easily find r$_\infty$.
 For the 2nd term of the potential, we will put the density of the fluid as the density of fluid in case of usual non self-gravitating spherically symmetric Bondi flow (The method of iteration discussed in section 5). There we need the density of the fluid at infinity. So, we will choose the density of fluid at infinity to be of such a small value so that the term $\frac{4\pi G}{r_\infty}\int_{R_\star}^{r_\infty}\rho r^{2}dr$ is negligible compared to the energy E of the fluid.\\
 Quantitatively, we set a very small quantity Q2 according to our desired precision as below.
 \begin{align*}
 Q2&=\frac{\frac{4\pi G}{r_\infty}\int_{R_\star}^{r_\infty}\rho r^{2}dr}{E}\\
 &=\frac{[\frac{4\pi G\rho_\infty (GM)^3I_{1\infty}}{r_\infty c_{s\infty}^6}]}{E}
 \end{align*} 
 \begin{equation}
 \rho_\infty=\frac{Q2 Er_\infty (E(Y-1))^3}{4\pi G(GM)^3I_{1\infty}}
 \end{equation}
 where we have used the method of iteration and as the 2nd and 3rd term of the equation (23) are taken to be negligibly small compared to the energy E of the infalling fluid we have used the following to find out c$_{s\infty}$.
 \begin{equation}
 E=\frac{c_{s\infty}^{2}}{(\gamma-1)}
 \end{equation}
 And also according to the method of iteration and using the value of r$_\infty$ from equation (24) we have,
\begin{equation}
I_{1c}=\int_{x*}^{x_b}z_{Bondi}(x)x^2dx
\end{equation}
\begin{equation}
I_{2c}=\int_{x_b}^{x_\infty}z_{Bondi}(x)xdx
\end{equation}
Now, we can numerically calculate the values $I_{1c}$ and $I_{2c}$ and put them in equation (11), (12) and (13) to find the changes in the critical parameters due to inclusion of fluid mass.
\section{Physical Importance of Q1 and Q2}
At infinity we can write the energy of infalling fluid as
\begin{align*}
E=\frac{c_{s\infty}^{2}}{(\gamma-1)}+EQ
\end{align*}
where $ Q (=Q1+Q2)$ gives total precision of the problem. So in summery, when E, M, $\gamma$ and Q are given, we at first set Q1 to find appropriate r$_{\infty}$ of the problem and setting Q2 accordingly gives us density of the infalling matter at infinity. If we closely look at the expressions of Q1 and Q2, we see that the ratio
\begin{equation}
s=\frac{Q2}{Q1}=\frac{\rm Mass~of ~the~ entire~ medium}{\rm Mass ~of~ the~ accretor}
\end{equation}
Where mass of the medium means the calculated mass of the medium considering density profile as that for usual Bondi flow, i.e; the mass of the medium calculated in the 1st iteration.
Intuitively, we can say that greater the value of ratio Q2/Q1 greater the effect of inclusion of fluid mass or greater the departure from the usual Bondi accretion.
\section{Behaviour of the critical parameters}
We can now find the critical parameters. We have the mass M of the accretor, E or the temperature of the fluid at infinity, $\gamma$ of the fluid as inputs and also we have fixed Q to be 0.02. Choosing s gives us r$_\infty$ and density of fluid at infinity. Our parameter space is $\gamma$ (4/3 to 5/3), E (0 to 1 in units of c$^{2}$). Now for neutral hydrogen-interstellar medium (H1 region)\cite{e} temperature varies nearly from 10$^{\circ}$K to 1000$^{\circ}$k and corresponding energy per unit mass varies nearly of the order of 10$^{-14}$ to 10$^{-12}$ in units of c$^{2}$ . Here we have plotted percentage change in the critical values due to inclusion of fluid mass with the $\gamma$-E parameter space. Temperature of the fluid at infinity is taken to vary nearly 50$^{\circ}$K to 1000$^{\circ}$K. 
We find the shift in transonic surface due to inclusion of fluid mass for different s.\\
\\
\begin{figure}
\includegraphics[scale=0.325]{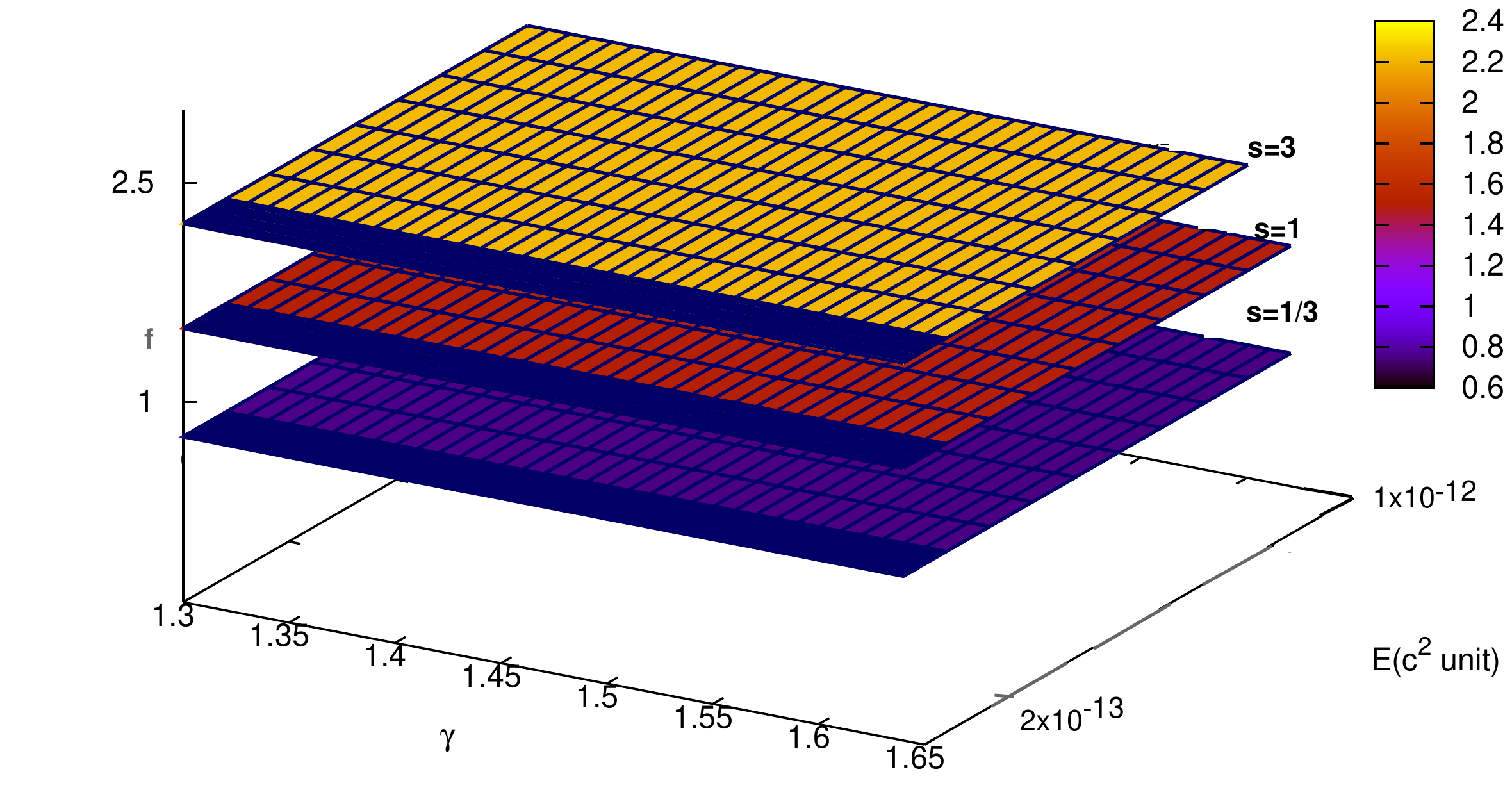} 
\\      
Fig. 1: Behaviour of percentage change in $x_{b}$ where f = -(Dx/x$_{b})\times$100 and Dx = (x$_{c}-x_{b}$)
\\
\end{figure}
The planes corresponding to different s seem parallel to $\gamma$-E plane because the variations in x$_{b}$ for different s are incomparable. The whole picture becomes clear when we plot for single s.\\
\\
\includegraphics[scale=0.32]{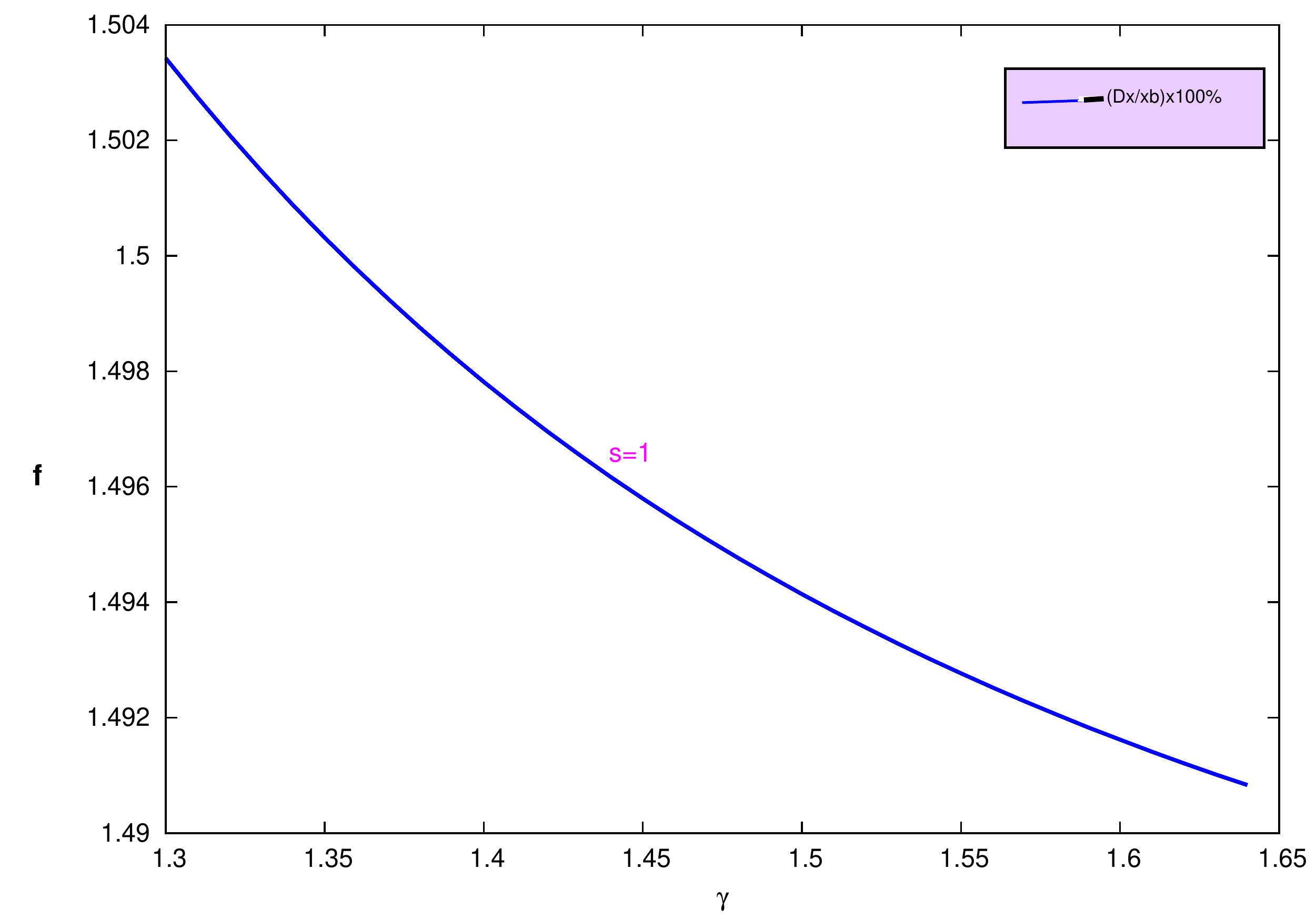}
Fig. 2: Projection of the surface of Fig. 1 for s=1 on (-Dx/x$_{b}$)x100-$\gamma$ plane.\\
\\
\includegraphics[scale=0.32]{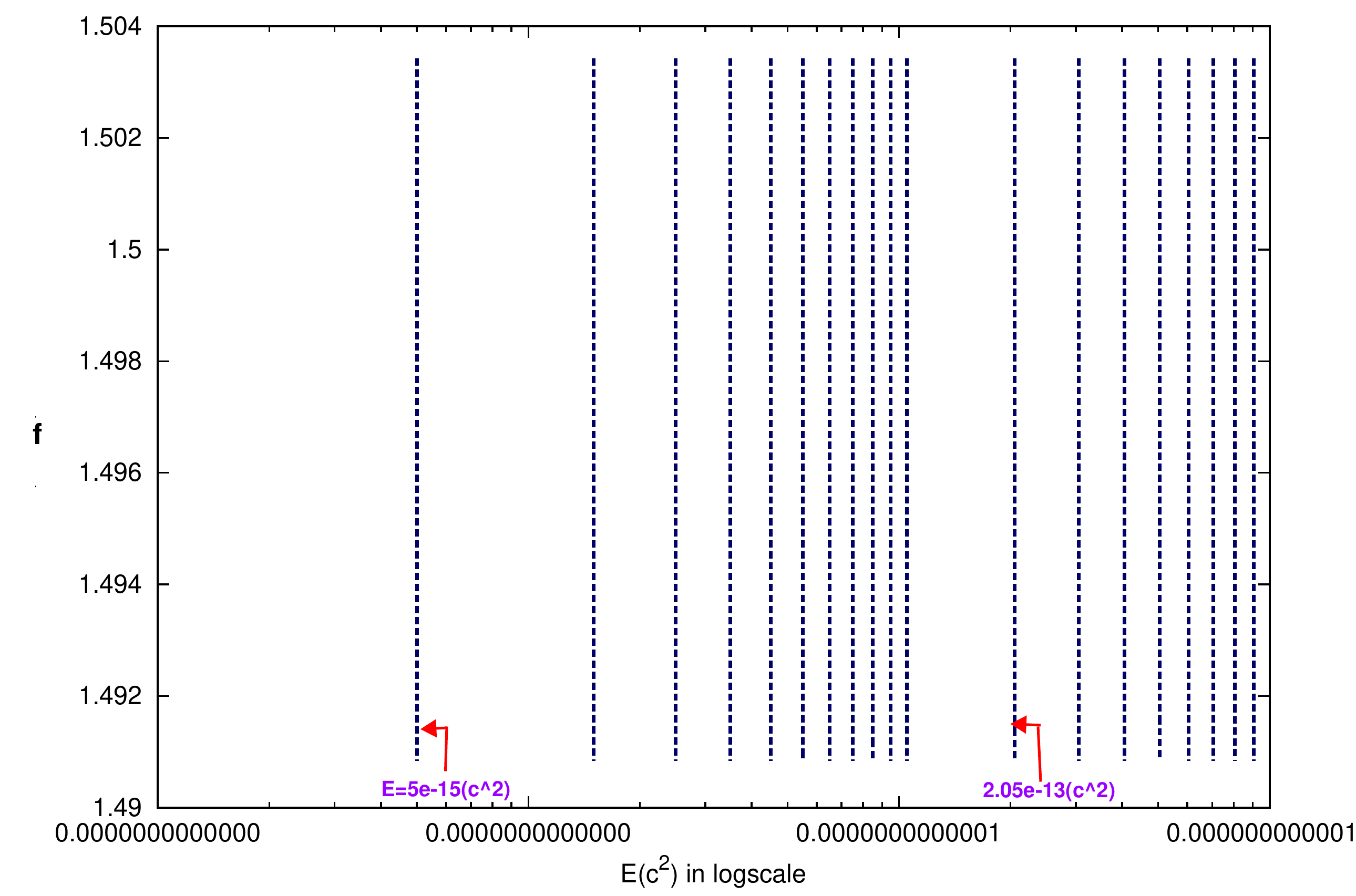}
Fig. 3: Projection of the surface of Fig. 1 for s=1 on (-Dx/x$_{b}$)x100-E (in log scale) plane.\\
Fig. 2 and Fig. 3 implies that when s and E are given and also total precision Q is given, the shift in transonic surface does not depend on E and that's why only a single curve appear in Fig. 2, otherwise there would be several curves for different Es. We can generate the surface for s=1 just by translation of the curve in Fig. 2 along E axis. The shift in critical point x$_{b}$ is same for all E for a particular $\gamma$ and Q. In the Fig. 3 the vertical lines signify the same thing, i.e; the points for a single vertical line gives the percentage change in x$_{b}$ for different $\gamma$s. A single vertical line corresponds to a single energy. Actually fixing Q and s of the problem give Q1 and Q1 in return gives r$_\infty$ for a given E. Then from Q2 deriving $\rho_{\infty}$ somehow wipe out the dependence on E. Inclusion of fluid mass decreases the radius of sonic surface More the total mass of the medium increases greater the shift in x$_{b}$. Relatively small $\gamma$s give significant changes.
We find the changes in other critical parameters.\\

\includegraphics[scale=0.33]{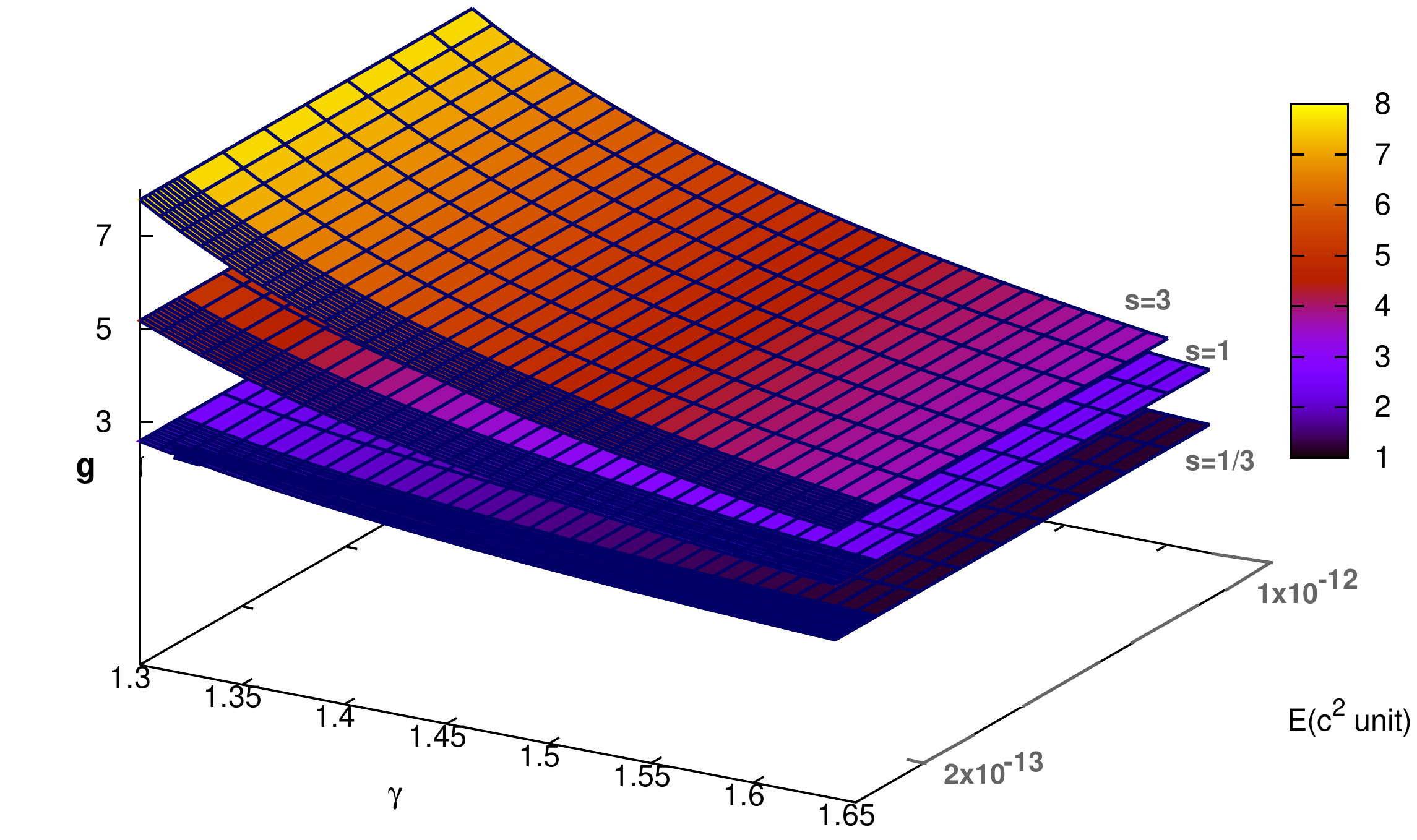}
Fig. 4: Behaviour of percentage change in $z_{b}$ where g = (Dz/z$_{b})\times$100 and Dz = (z$_{c}-z_{b}$)\\

\includegraphics[scale=0.32]{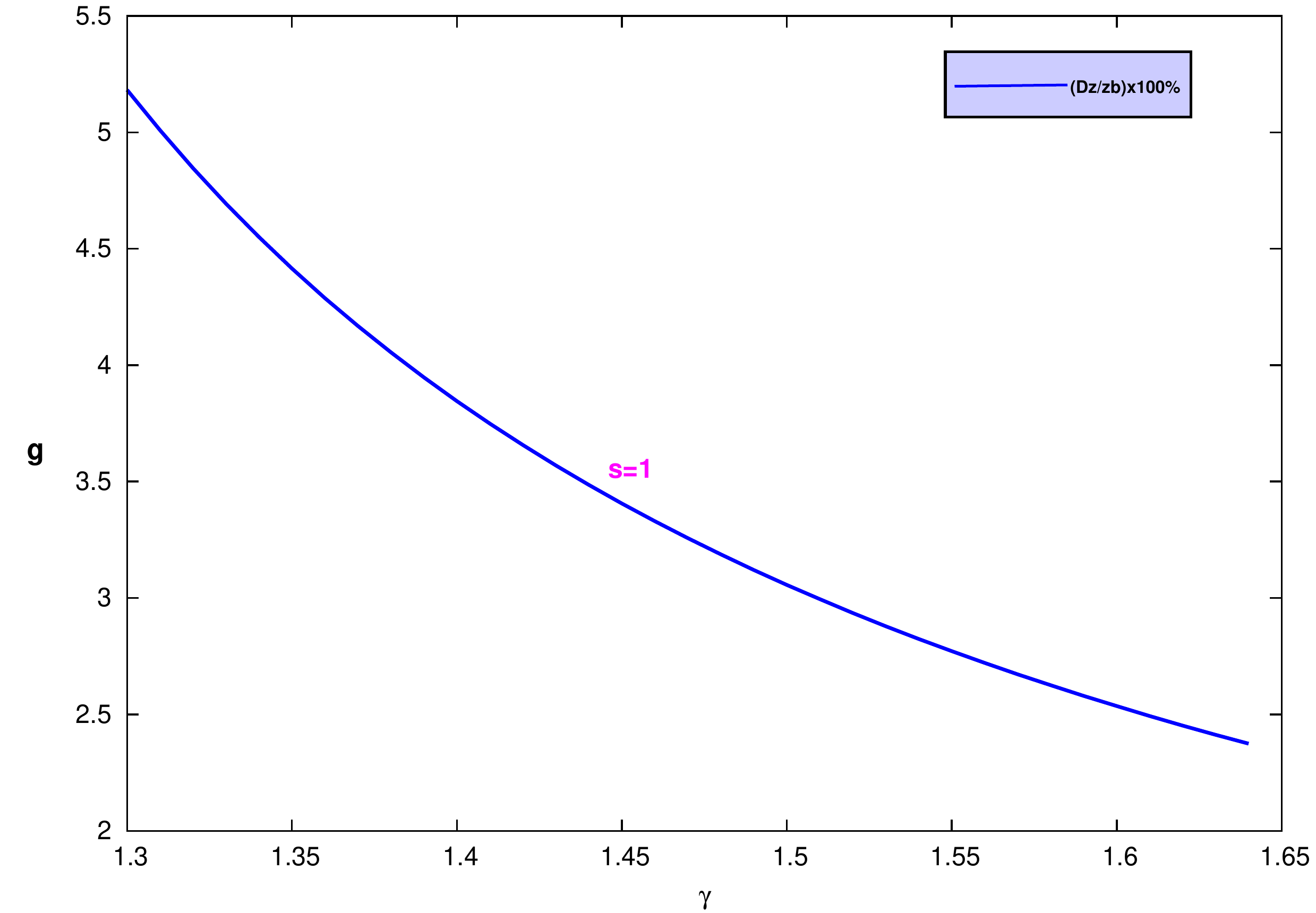}

Fig. 5: Projection of the surface of Fig. 4 for s=1 on (Dz/z$_{b}$)x100-$\gamma$ plane.\\
\\
For brevity, we have not plotted E-(Dz/z$_{b}$)x100. Similarly, percentage change in z$_{b}$ does not depend on E and z$_{b}$ is significantly increased for $\gamma$ close to 4/3. Fig. 4 implies that if we increase s we increase the change in z$_{b}$ which is intuitively obvious. Another important point is that if we compare the percentage changes in z$_{b}$ with that of x$_{b}$ we see that inclusion fluid mass effect z$_{b}$ more than x$_{b}$.\\
\\
\\
\includegraphics[scale=0.325]{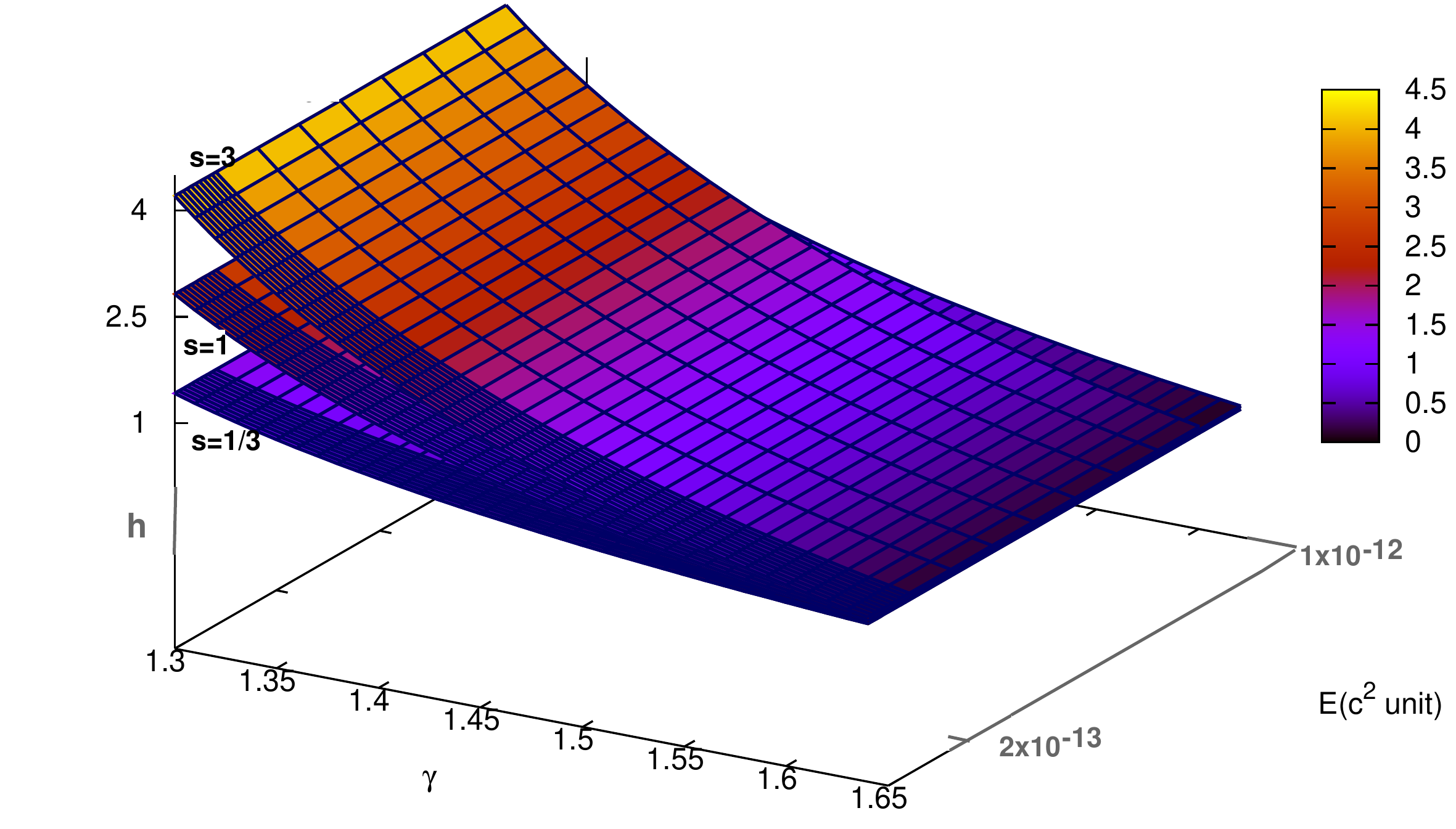}
\\
\\
Fig. 6: Behaviour of percentage change in $\lambda_{b}$ where h $=(D\lambda/\lambda_{b})\times$100 and D$\lambda=(\lambda_{c}-\lambda_{b}$)\\
The surfaces seem to converge towards $\gamma$=1.65.\\
\\
\\
\includegraphics[scale=0.325]{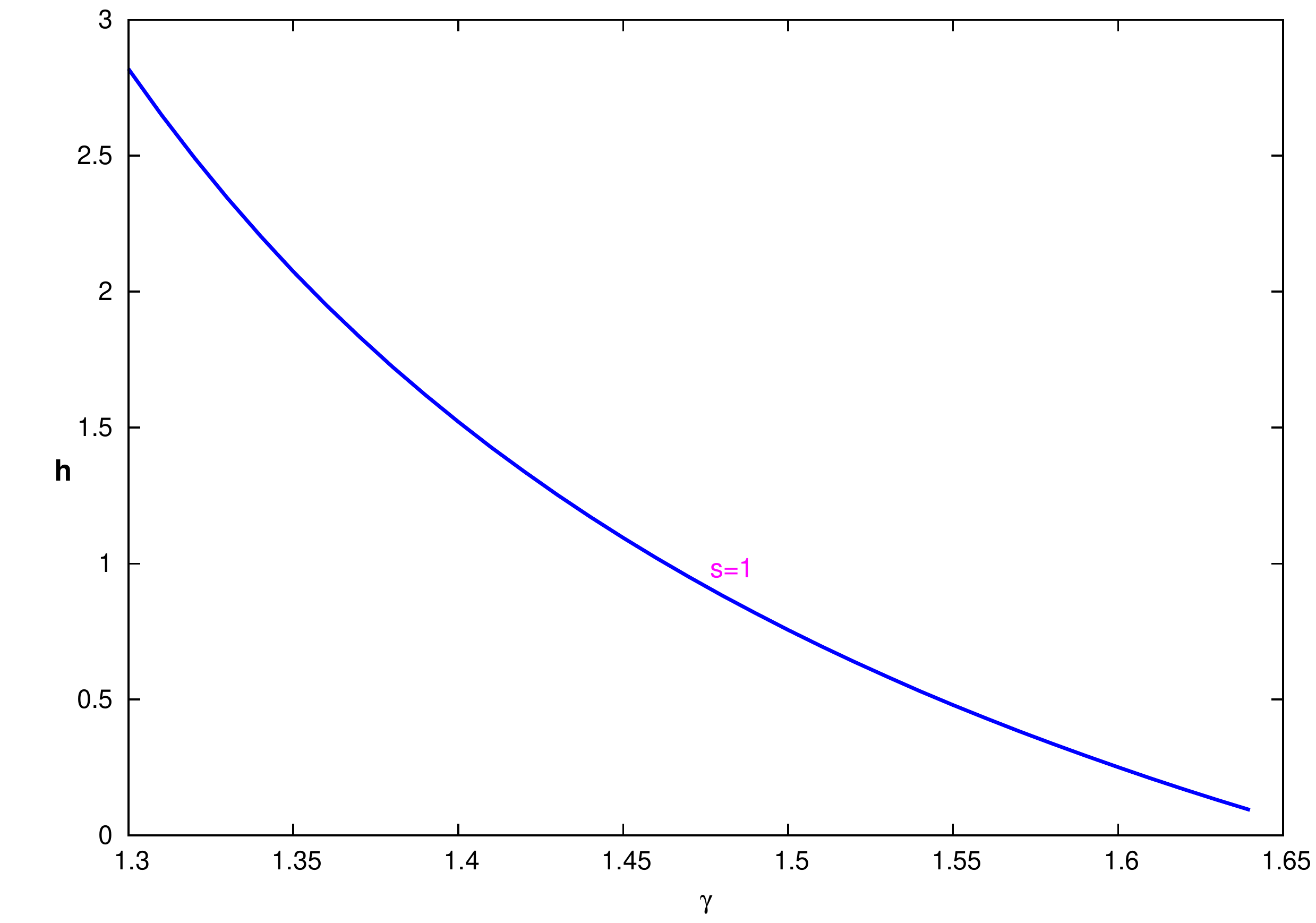}
Fig. 7: Projection of the surface of Fig. 6 for s=1 on (D$\lambda/\lambda_{b}$)x100-$\gamma$ plane.\\
\\
Similarly, percentage change in $\lambda_{b}$ does not depend on E and it is significantly increased for $\gamma$ close to 4/3. Observing Fig. 6 implies that if we increase s we increase the change in $\lambda_{b}$ which is intuitively obvious.\\
Increasing s increases the shift in the critical parameters.\\ 
Another important observation is that the percentage change in x$_{b}$ with respect to $\gamma$ is not only smaller than the other critical parameters but also the variation in percentage change of x$_{b}$ with $\gamma$ is negligible compared to that for the other critical parameters. One can safely say that percentage change in x$_{b}$ merely depends on $\gamma$ and actually it's a function of only one variable, s. Again s can not be arbitrarily large because in that case the change in mass of the accretor will be significant. Inclusion of fluid mass changes z$_{b}$ much more significantly than the other critical parameters. We find how do the critical parameters depend on the mass ratio of the fluid medium and the star.
\\
\\
\includegraphics[scale=0.32]{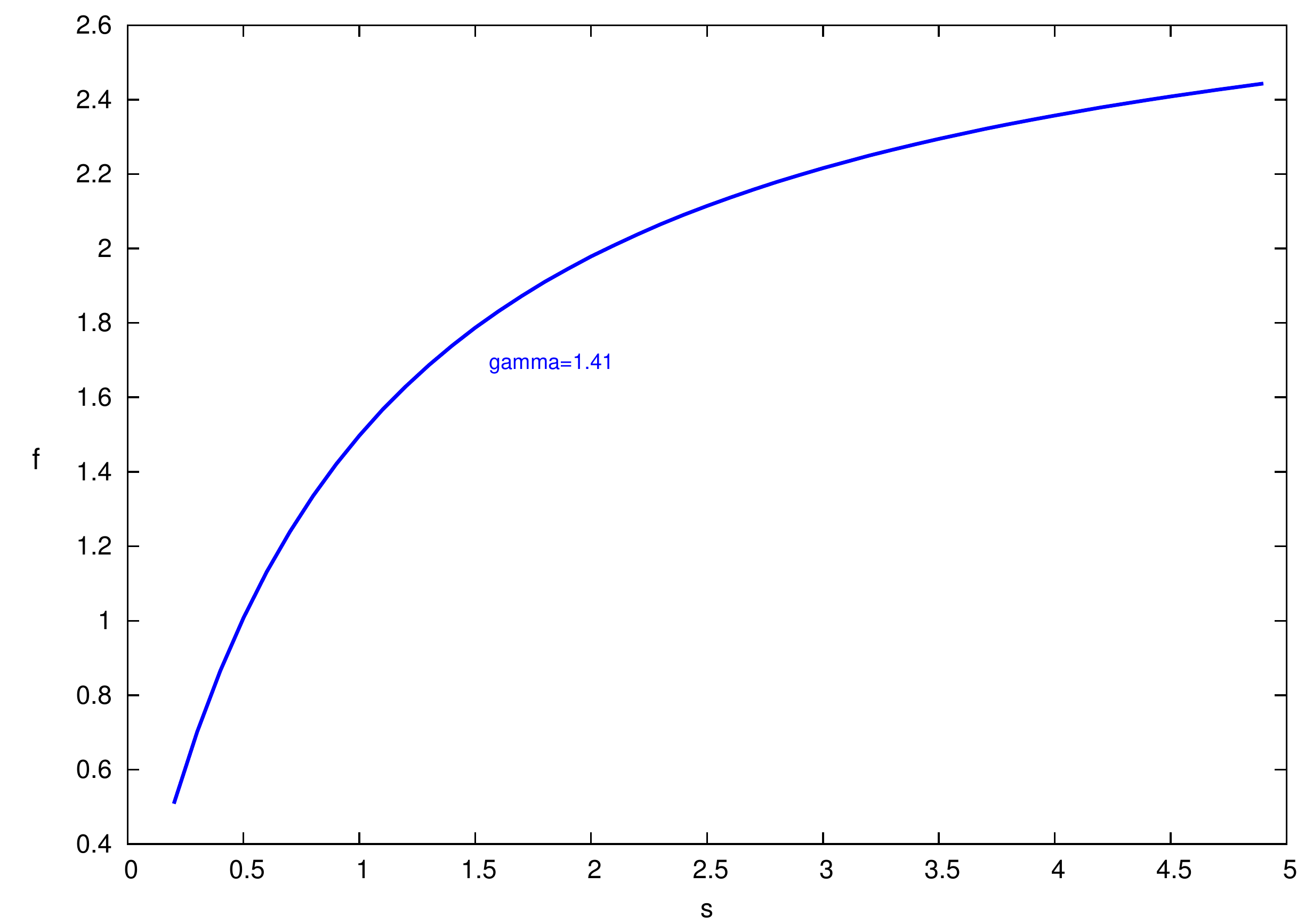}\\\\\includegraphics[scale=0.32]{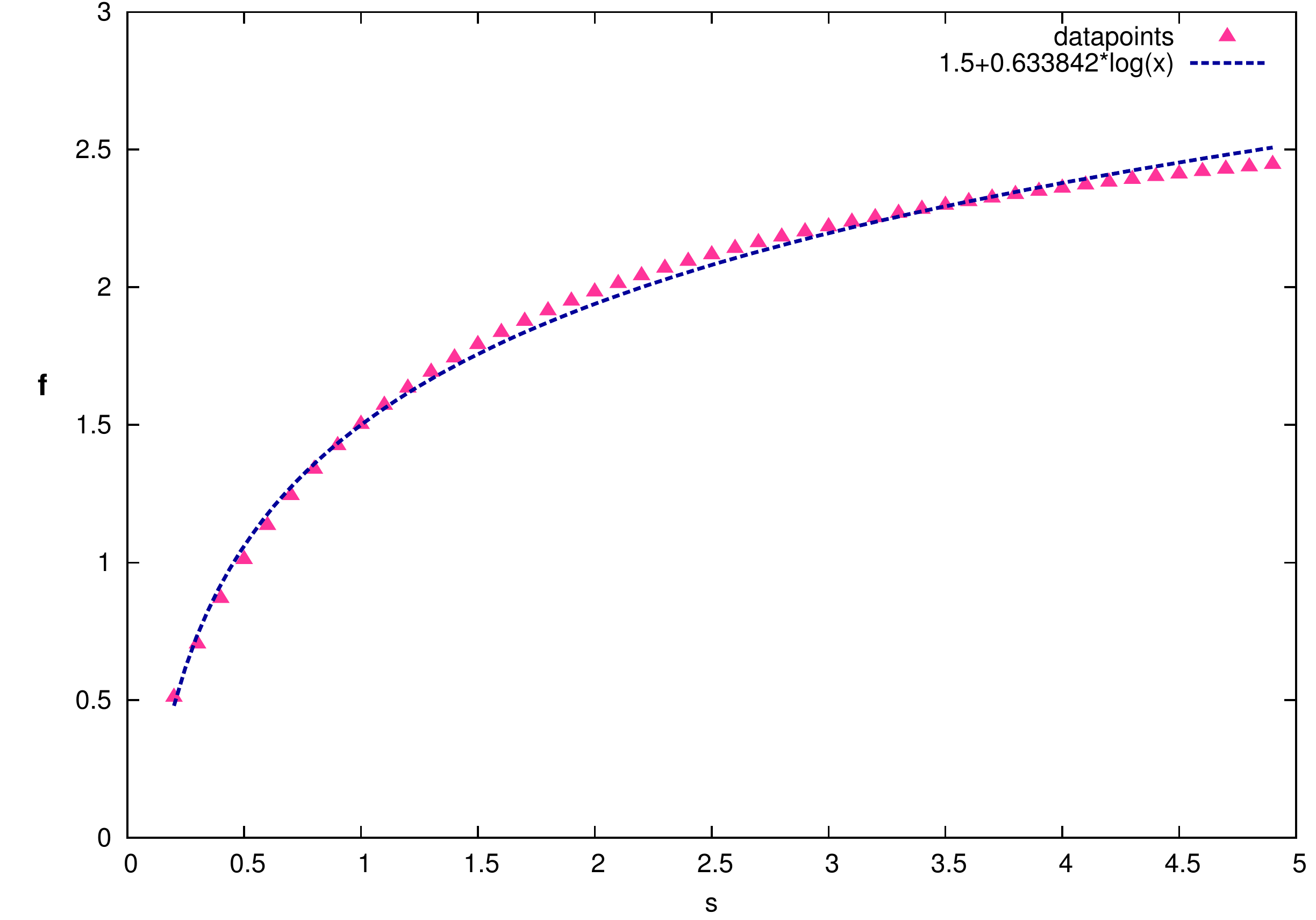}\\
Fig. 8: Dependence of percentage change in x$_{b}$ on s and data points are fitted. We choose upper limit of s to be 5.\\
\\
As percentage change in x$_{b}$ is only function of s, we have fitted the data points to find the nature of dependence on s. We have obtained an empirical relation as
\begin{equation}
f=(-\frac{Dx}{x_{b}}\times 100)=a+b ~ln(s)
\end{equation}
In our case, the positive constants are as a = 1.5 and b = 0.633842. The fact that a to be 1.5 is also obvious from Fig. 2.\\
\\
\includegraphics[scale=0.32]{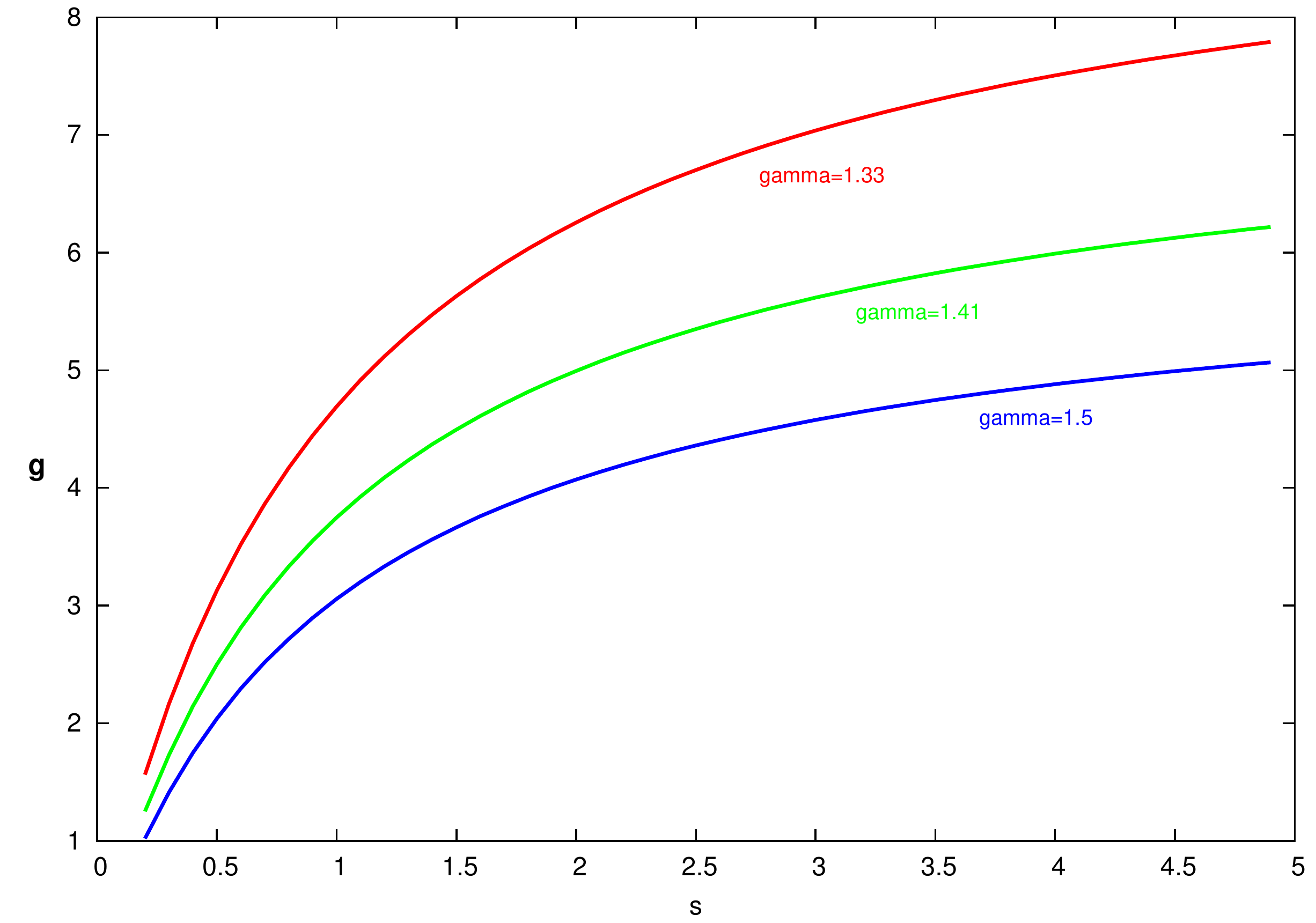}
Fig. 9: Dependence of percentage change in z$_{b}$ on s.\\
\\
\includegraphics[scale=0.32]{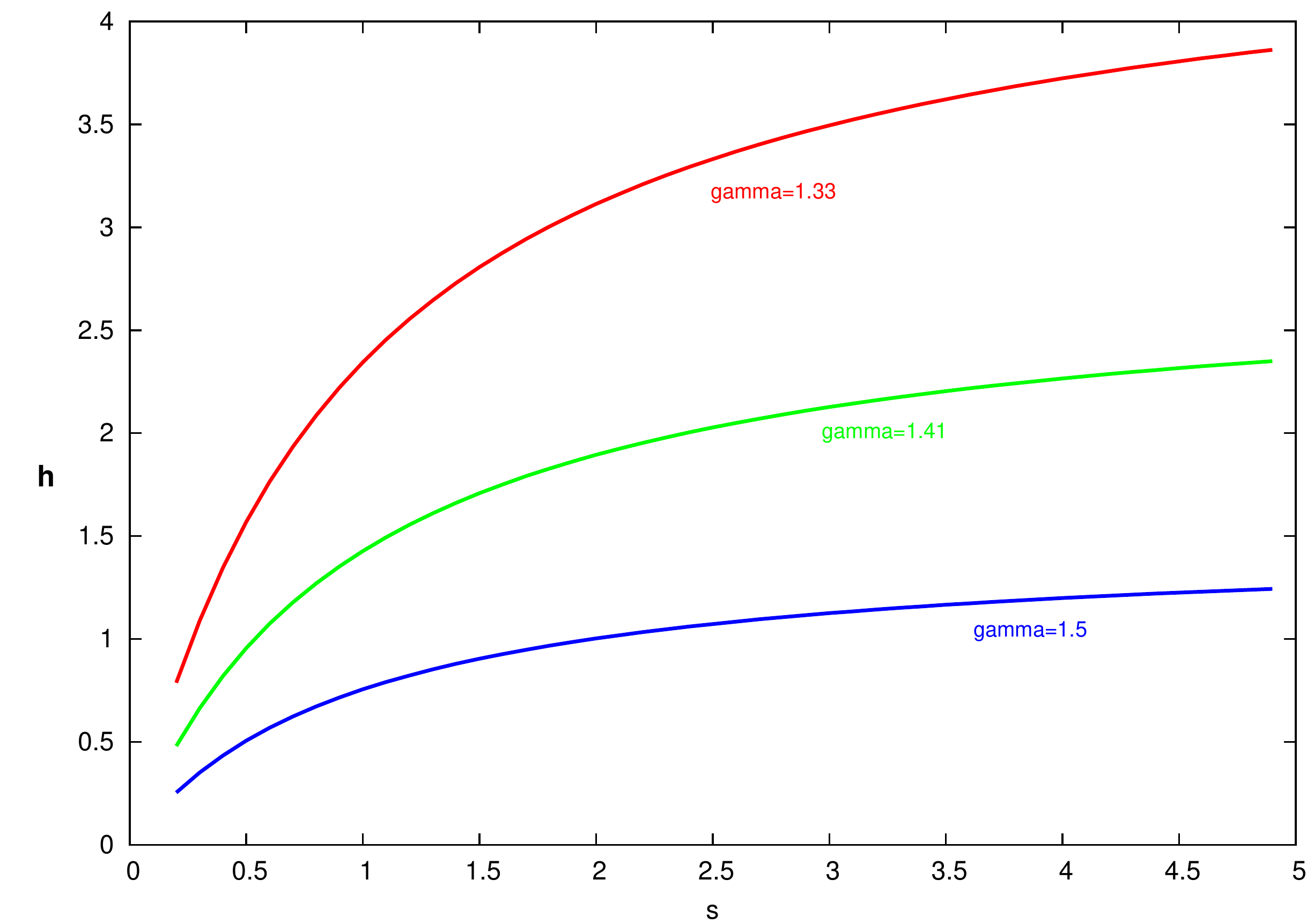}
\\
Fig. 10: Dependence of percentage change in $\lambda_{b}$ on s.\\
\\
The changes in critical parameters increase with s. Fig. 9 and Fig 10. shows also that why do the surfaces for z$_{b}$ and $\lambda_{b}$ converge towards $\gamma$ = 1.65. We see that the separations between the curves of Fig. 9 and 10 for $\gamma$ = 1.33 and 1.41 is greater than the separation between the curves for $\gamma$ =1.41 and 1.5. \\
Equation (30) implies that the test fluid approximation is good for s $\leqslant$ exp(-a/b). This is the region of s where Bondi's assumptions and his results are applicable. In our case, that region corresponds to s $\leqslant$ 0.094 and it's important to mention that  a and b depends on Q. Similarly one can fit the same for $\lambda_{b}$ and z$_{b}$ as well and one will find some best fitting curves as (f$_{\lambda, z}$($\gamma$)+g$_{\lambda, z}$($\gamma$)ln(s)) respectively. From Fig. 7 and 5 we can conclude that f$_{\lambda, z}$($\gamma$) are monotonically decreasing functions of $\gamma$. As the curves for other s have same nature as s=1, g$_{\lambda, z}$($\gamma$) can not be monotonically increasing function and as the surfaces are not parallel to each other which is obvious from Fig. 6 and 4, g$_{\lambda, z}$($\gamma$) is also monotonically decreasing function of $\gamma$.\\
Now we find find $\alpha$\cite{b} dependence of the percentage change in critical parameters. \\
\\
\includegraphics[scale=0.32]{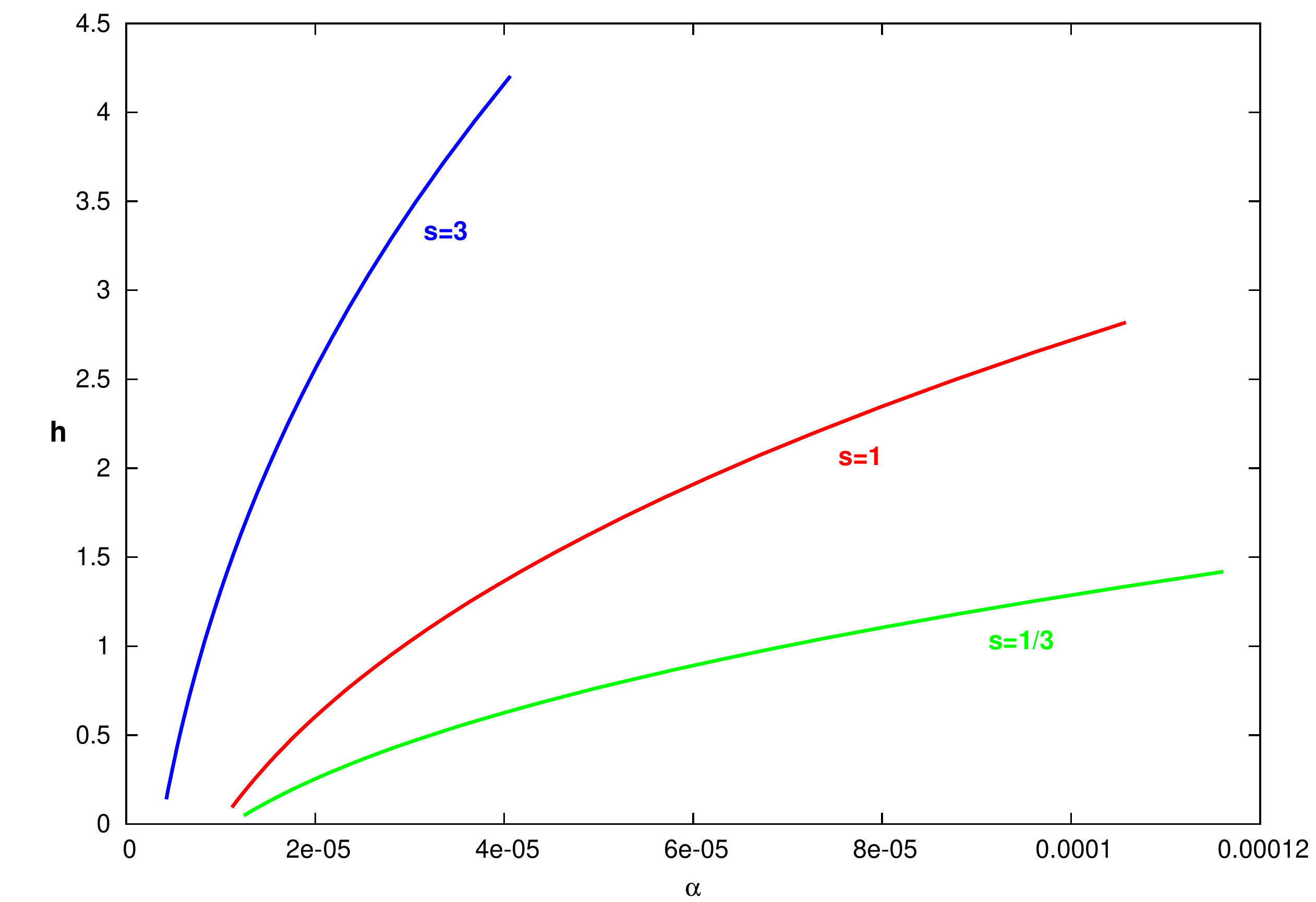} 
\\
Fig. 11: Dependence of percentage change in $\lambda_{b}$ on $\alpha$.\\
\\
For brevity, we have not plotted the same for x$_{b}$ and z$_{b}$ because they are more or less of same nature. Increasing $\alpha$ increases percentage change in the critical parameters. Again as percentage change in x$_{b}$ is a slowly varying function of $\gamma$, it is also a slowly varying function of $\alpha$. We have fitted the above plot as below,\\
\\
\includegraphics[scale=0.32]{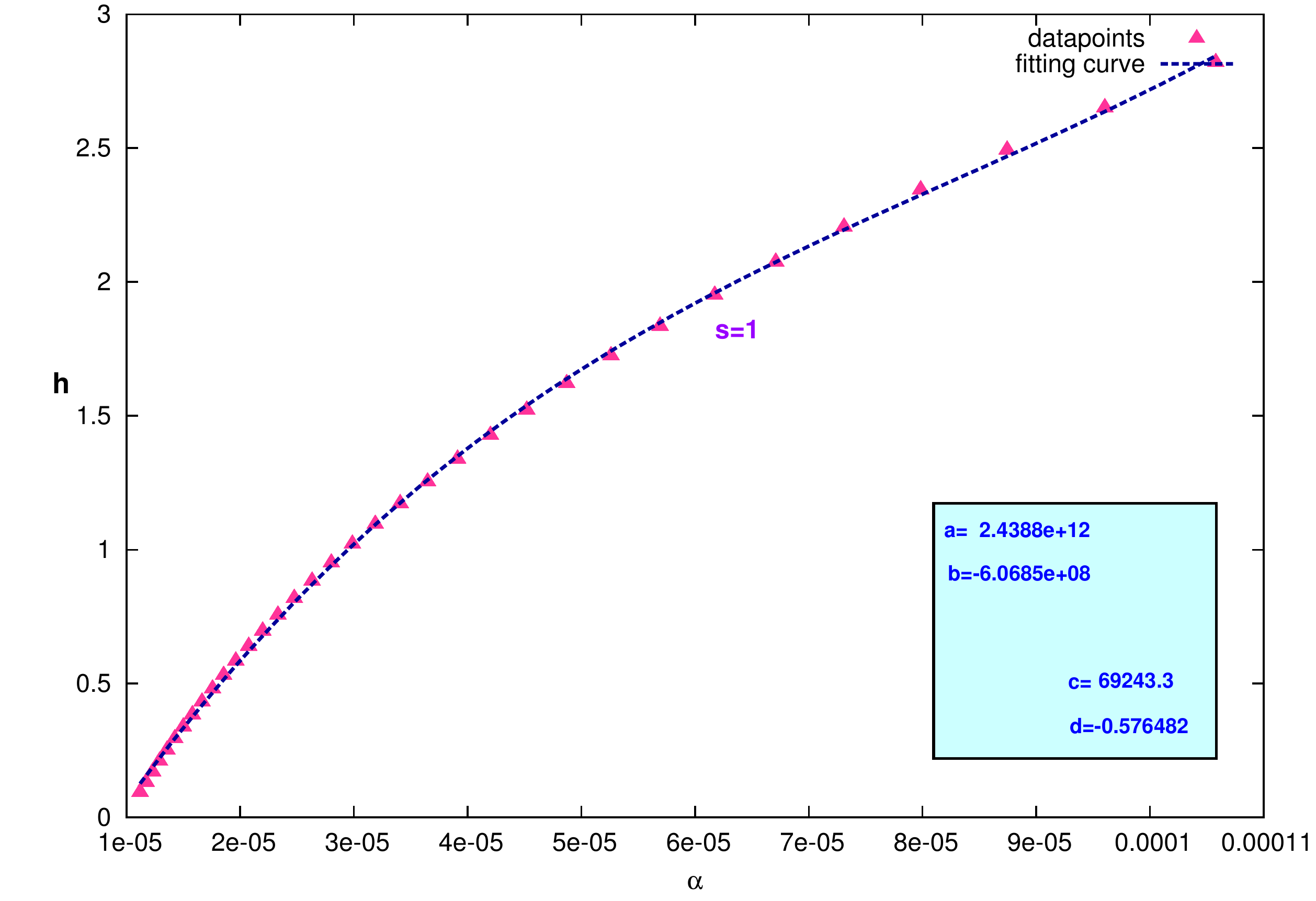}
\\
Fig. 12: The data points are fitted to the curve a$\alpha^{3}$+b$\alpha^{2}$+c$\alpha$+d\\
\\
We have found empirically a cubic polynomial,
\begin{equation}
h=(\frac{D\lambda}{\lambda_{b}})\times100=a\alpha^{3}+b\alpha^{2}+c\alpha+d
\end{equation}
The coefficients a, b, c, d depend on s, E and Q. Similarly percentage change in z$_{b}$ and x$_{b}$ have cubic dependence on $\alpha$ and in the same way, we can fit them into cubic polynomials (not shown for brevity).
\section{Ambiguity of the velocity gradient}
Now, we can calculate du/dr at transonic surface from equation (18). We have obtained the critical parameters, we will now just put them to find q. Important point is to note that in $I_{1c}$, appearing in the term C of the quadratic equation of q, we will use the new value of x$_c$ which is found by 1st iteration. Now, according to equation (18), we will obtain two values of q. As inclusion of fluid mass does not changes the flow abruptly so we will choose the value of q which is closer to the value of critical (du/dr) when fluid mass effect isn't considered. So, we plot the behaviour of the derivatives and have compared those values with the spherical Bondi flow without including fluid mass. We are calling the solutions of q as below,
\begin{align*}
& q_1=\frac{-B+\sqrt{B^2-4AC}}{2A}\\
& q_2=\frac{-B-\sqrt{B^2-4AC}}{2A}
\end{align*}
and we are calling
\begin{align*}
q=\frac{du}{dr}|_{Bondi}
\end{align*}
where $\frac{du}{dr}|_{Bondi}$ is the value of du/dr at transonic surface without inclusion of gravity due to medium.\\
\\
\includegraphics[scale=0.32]{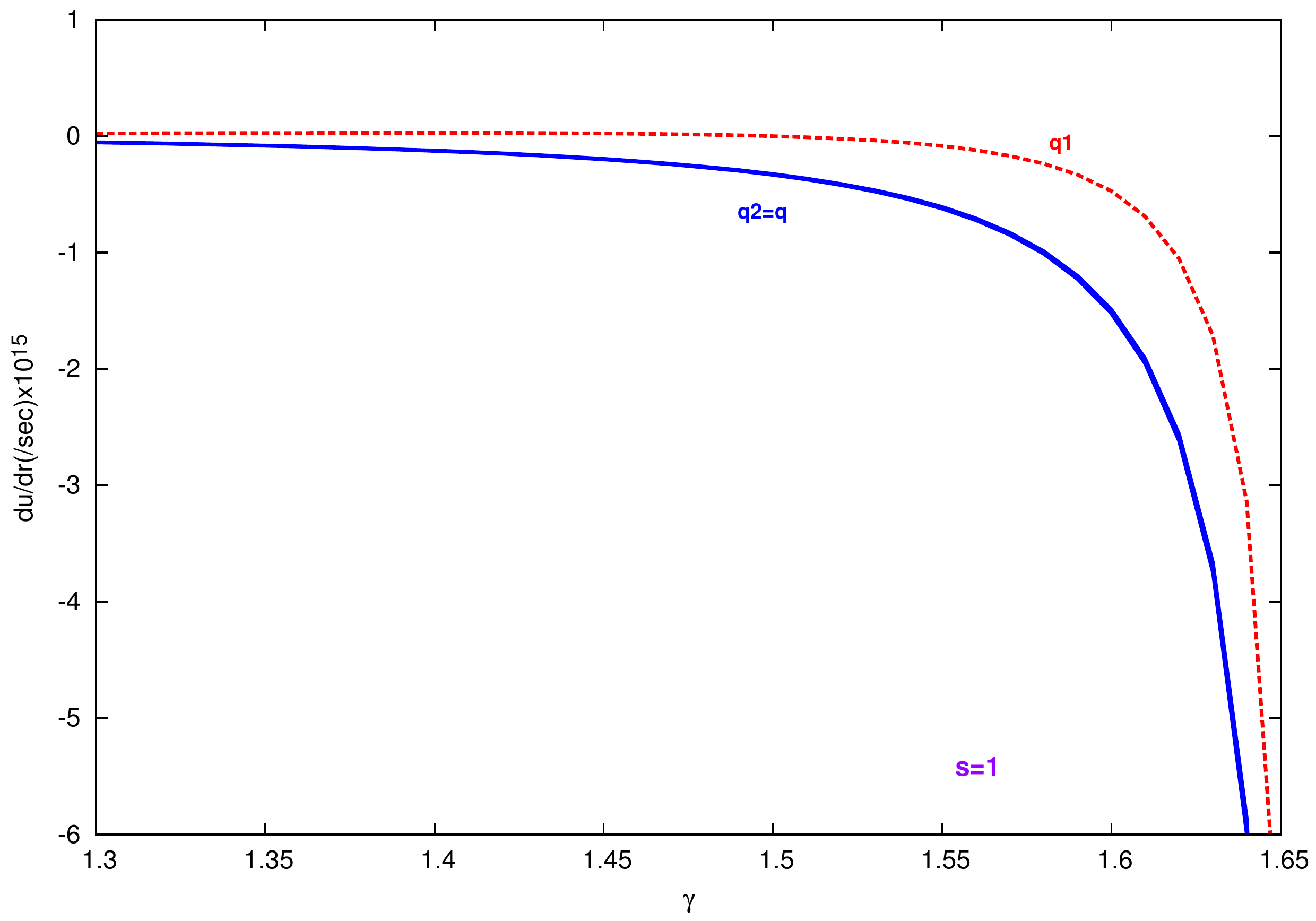}
Fig. 13: q$_1$, q$_{2}$ and q are plotted against $\gamma$ for comparison\\
From the above figure, we see that q$_2$ is much closer to q than q$_1$. As we are considering  that the inclusion of fluid mass does not alter the critical values abruptly, so we will choose q$_2$ over q$_1$.
\section{Mach number vs radial distance profile}
Now, we have found the initial conditions to find variation of Mach number with radial distance from the centre of the accretor. To do this, we have to numerically solve equation (16) and (17) simultaneously. Mach number(=u/c$_s$) can then be found. We have now x$_c$, y$_c$, z$_c$ and q$_{2}$ as initial conditions and we will use the numerical values of density for spherically symmetric Bondi flow without including fluid mass effect. So we find Mach number vs radial distance profile as below.
\includegraphics[scale=0.3]{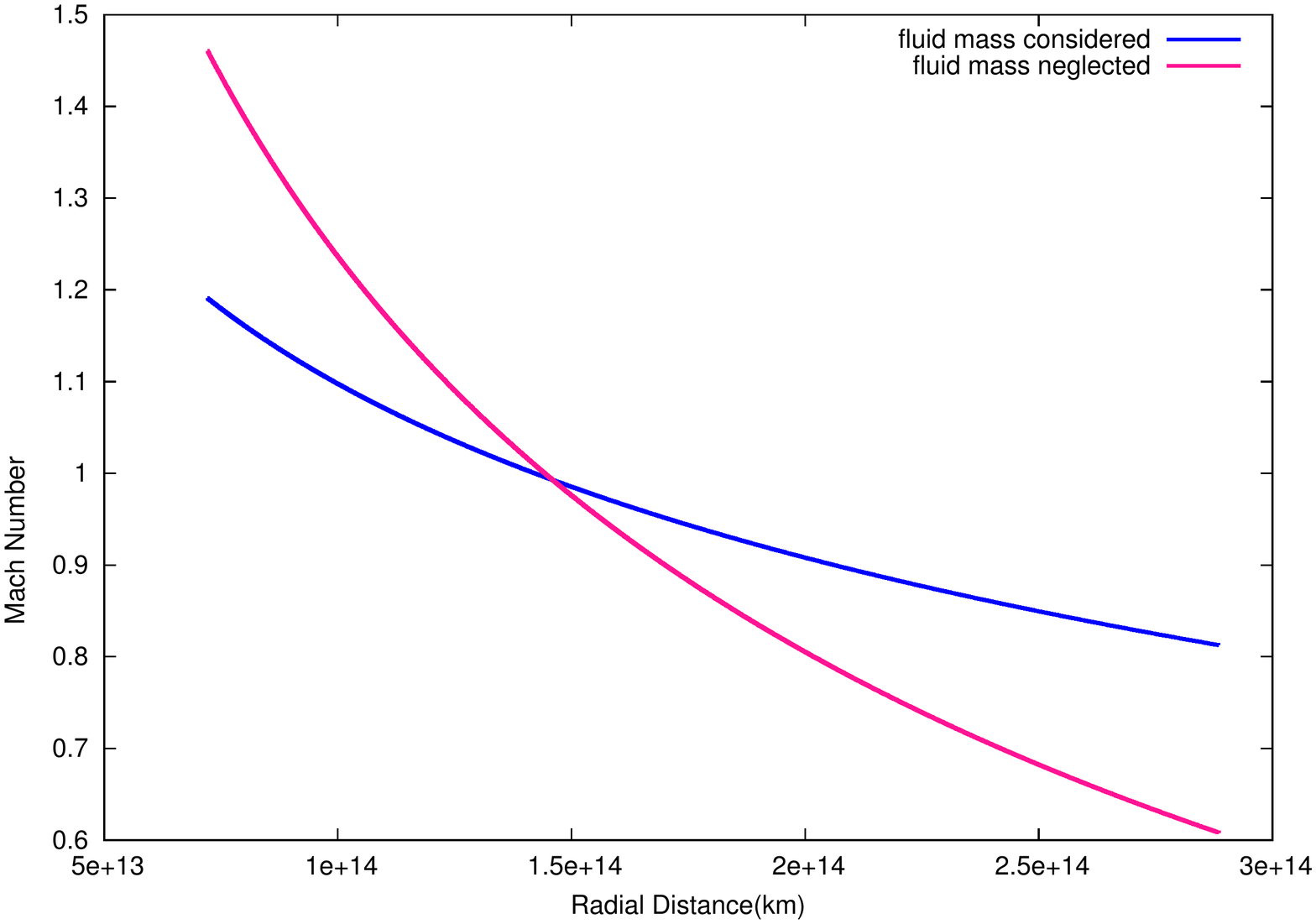}
Fig. 14: Mach number vs radial distance profile where mass of the Main-Sequence star is 1 solar mass and $\gamma$ of the fluid is 1.41, temperature of the infalling fluid at infinity is about 50$^{\circ}$K and density of the fluid at infinity is of the order 10$^{-29}$kg/m$^3$.\\
\\
\includegraphics[scale=0.3]{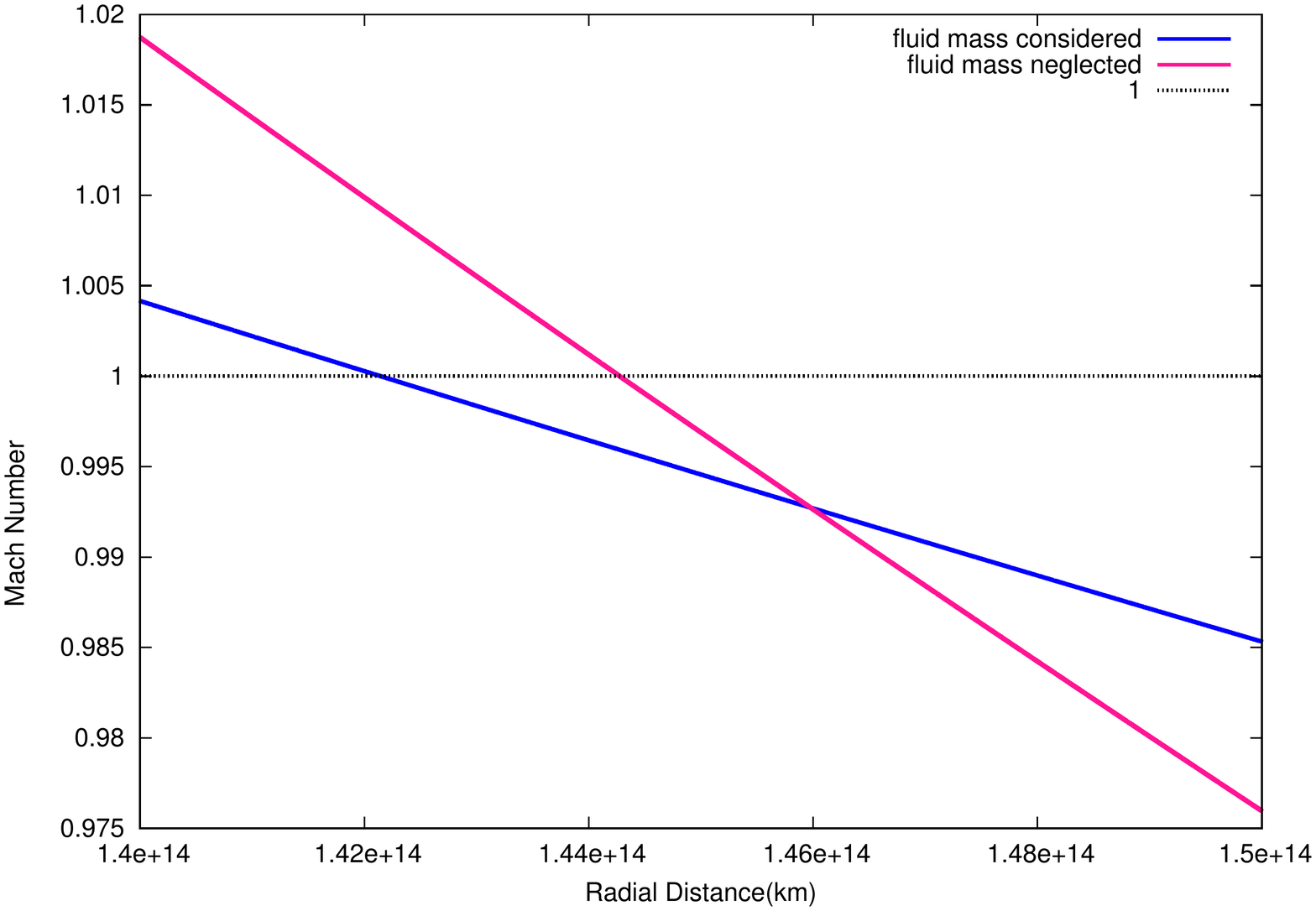}
Fig. 15: Close view near transonic surface.\\
\\
So, inclusion of fluid mass make the infalling fluid to cross sound speed nearer to the accretor.
\\
\section{Summary and Conclusions}
In this paper, we have studied the non relativistic spherically symmetric accretion including fluid mass. We haven't considered the growth of the accretor during the process of accretion. We have considered steady flow of infalling fluid and the accretor to be a candidate of Main-sequence stars. \\We have designed a methodology to find the changes in spherically symmetric accretion when fluid mass is taken into account. We have set Q, i.e; the precision of the problem and then we have considered several accretion systems by our input parameters. We may have in general several inputs as: mass M of the accretor, $\gamma$ of the fluid, energy E of the fluid, density of the fluid at infinity $\rho_{\infty}$, the mass ratio of the medium and the accretor s and the extent of the medium r$_{\infty}$ etc. Now according to our methodology after fixing Q when we take s into account that in return fixes the extent of the medium r$_{\infty}$ according to the input energy E. $\rho_{\infty}$ is then fixed in accordance with $\gamma$ and E of the fluid. In a summary when we have given the 5 input parameters M, Q, s, $\gamma$ and E of the problem r$_{\infty}$ and $\rho_{\infty}$ are fixed. r$_\infty$ and $\rho_{\infty}$  are no longer independent input parameters. Actually, these 5 input parameters contain the informations about $\rho_{\infty}$ and r$_{\infty}$. One can start in other way round like by taking input parameters, r$_{\infty}$ and $\rho_{\infty}$ first and then setting Q of the problem. In that case, our 5 independent parameters would be M, Q, $\gamma$, r$_{\infty}$ and $\rho_{\infty}$. Then one can find the changes in the critical parameters by varying r$_{\infty}$ and $\rho_{\infty}$. That will be another way of looking at the same problem. Similarly, one can have other several ways of looking at the same problem. \\
Now if an arbitrary accretion system is given in a way that some input parameters are known then at first we have to check that the known input parameters are sufficient to find any solution or not and if the given input parameters are sufficient to analyse the system then we have to find what is the precision Q and if Q is close to 1 then the system is outside of our formalism. Otherwise, we will try to find s of the problem and if s is too small or if s lies inside the region where test fluid approximation is valid (as discussed in section IX) then one can safely proceed without including fluid mass effect otherwise one have to proceed by following the procedure discussed in the above sections.\\
 Using this methodology, we have compared the results with the usual spherically symmetric Bondi flow and we have found that inclusion of fluid mass changes the nature of flow of the infalling fluid. Mass accretion rate is increased  and radius of the transonic surface is decreased when fluid mass comes into picture.
 \section{Acknowledgements}
 The author is thankful to his Ph.D. supervisor for introducing him 
 to this problem and for his help in modifying the calculations and 
 the overall presentation. He is also thankful to his colleague Dhruv Pathak
 for useful discussions. 

\end{document}